\documentclass[a4paper,fleqn,usenatbib]{mnras}
\usepackage{amssymb,amsmath,graphicx,psfrag,mathtools}
\usepackage[T1]{fontenc} 
\usepackage{ae,aecompl} 
\usepackage{url}
\usepackage{revsymb}
\hypersetup{draft} 
\title[FRB 121102 and SGR/PSR J1745$-$2900]{The Environment of FRB 121102 and Possible Relation to SGR/PSR J1745-2900}
\author[J. I. Katz]{
J. I. Katz,$^{1}$\thanks{E-mail katz@wuphys.wustl.edu} 
\\
$^{1}$Department of Physics and McDonnell Center for the Space Sciences,
Washington University, St. Louis, Mo. 63130 USA 
}
\date{Accepted XXX.  Received YYY; in original form ZZZ} 
\pubyear{2020} 
\date{\today}
\begin{document} 
\label{firstpage} 
\pagerange{\pageref{firstpage}--\pageref{lastpage}} 
\maketitle 
\begin{abstract}
	Variations of the dispersion (DM) and rotation (RM) measures of FRB
	121102 indicate magnetic fields $\sim$ 3--17 mG in the dispersing
	plasma.  The electron density may be $\sim 10^4\,$cm$^{-3}$.  The
	observed time scales $\sim 1$ year constrain the size of the plasma
	cloud.  Increasing DM excludes simple models involving an expanding
	supernova remnant, and the non-zero RM excludes spherical symmetry.
	The varying DM and RM may be attributable to the motion of plasma
	into or out of the line of sight to or changing electron density
	within slower-moving plasma.  The extraordinarily large RM of FRB
	121102 implies an environment, and possibly also a formation process
	and source, qualitatively different from those of other FRB.  The
	comparable and comparably varying RM of SGR/PSR J1745$-$2900
	suggests it as a FRB candidate.  An appendix discusses the age of
	FRB 121102 in the context of a ``Copernican Principle''.
\end{abstract}
\section{Introduction}
\citet{WWY} recently reviewed the magnetoionic environments of Fast Radio
Bursts (FRB).  The dispersion measures (DM) and rotation measures (RM) of
most FRBs have shown no detectable changes and are consistent with paths
through the intergalactic medium and the interstellar media of our Galaxy
and a host galaxy (whose properties are necessarily very uncertain).  There
is one striking exception, FRB 121102, whose RM has extraordinary values
$\sim 10^5$/m$^2$, decreased by $\approx 30\%$ during about three years of
observations, and whose DM increased by about 2.5 pc/cm$^3$ \citep{H20}.
From these data it is possible to estimate the magnetic field and electron
density in the dense, strongly magnetized, region in which the RM and the
varying part of the DM are produced.  No other FRB has such a high RM, but
the other phenomenology of FRB 121102 is not extraordinary among FRB.

The magnetoionic environment of SGR/PSR J1745$-$2900 may be similar to that
of FRB 121102; its RM is also large and rapidly varying, although no change
in its DM has been reported.  If these two objects are related, SGR/PSR
J1745$-$2900 would be a candidate FRB source.
\section{Magnetic Fields}
The mean parallel component of magnetic field along the line of sight is
related to the RM and DM:
\begin{equation}
	\label{B}
	\langle B_\parallel \rangle_{n_e} = 1.23 {\text{RM} \over \text{DM}}
	\mu\text{G},
\end{equation}
where the RM is in units of radians/m$^2$, the DM is in units of pc/cm$^{3}$
and the subscript $n_e$ indicates this is the electron density-weighted
average of the parallel (to the line of sight) component of magnetic field.
In a homogeneous medium $\langle B_\parallel \rangle_{n_e}$ equals a single
value $B_\parallel$.

The fact that the RM is an integral of the product $B_\parallel n_e$ implies
that it is dominated by the densest and most strongly magnetized regions,
even if they make only a small contribution to the DM; the RM contains two
powers of parameters that are expected to be correlated, while the DM is
linear.  Most of the DM is likely produced in long paths in low $B$
interstellar and intergalactic plasmas that contribute negligibly to the RM.
In order to determine the magnitude of $B_\parallel$ it is necessary to
estimate the DM of the regions that dominate the RM.  Only then can
Eq.~\ref{B} be used to estimate $\langle B_\parallel \rangle_{n_e}$.
Unfortunately, for most FRB it is not possible to separate any contribution
to the DM of a dense and strongly magnetized region from those of the
interstellar (in both galaxies) and intergalactic media.

FRB 121102 is an exception.  Not only is its RM extraordinarily large and
rapidly varying, but it is the only FRB whose DM is known to change
\citep{H20,O20}, increasing by about 2.5 pc/cm$^3$ in two years of accurate
measurements.  Characteristic time scales of variation of contributions
to the DM are $T \sim L/v$, where $L$ is a length scale and $v$ a velocity.
For interstellar media $T \sim \text{100 pc}/(\text{30 km/s}) \sim 3 \times
10^6\,$y, while for the intergalactic medium $T \sim 1/h \sim 10^{10}\,$y,
where $h$ is the Hubble constant; the corresponding $d\text{DM}/dt \sim
\text{DM}/T$ are unobservably small ($\ll 10^{-3}\,$pc/cm$^3$ during the
period of observation).  The observed change in DM must be attributed to a
compact dense region that is also the source of most of the (also varying)
RM.

It is then possible to estimate the magnitude of $B_\parallel$ in this
region.  In the simplest possible model the dense region is homogeneous, is
the source of the entire RM, and its RM and DM vary in proportion.  If
correct, this would also permit estimating this region's contribution to the
DM.  A possible physical realization would be the intrusion into, or the
withdrawal from, the line of sight of a homogeneous (in density and magnetic
field) wedge of plasma, or a change in its ionization state.  Then
Eq.~\ref{B} becomes \citep{K18}
\begin{equation}
	\label{DB}
	\left|\langle B_\parallel \rangle_{n_e}\right| = 1.23{|\Delta
	\text{RM}| \over |\Delta \text{DM}|}\,\mu\text{G}.
\end{equation}

The observations of FRB 121102 by \citet{H20} fall into three groups: I,
comprising bursts 5 and 6 during MJD 58069--58075; II, comprising bursts
7--15 during MJD 58216--58348; and III, comprising bursts 16--20 during MJD
58678--58712.  Between groups I and II $\Delta \text{RM} \approx -15000$
m$^{-2}$ and $\Delta \text{DM} \approx 1.1$ pc/cm$^3$; between groups II and
III $\Delta \text{RM} \approx -4000$ m$^{-2}$ and $\Delta \text{DM} \approx
1.6$ pc/cm$^{-3}$.  Groups I and II are separated by about 160 days while
groups II and III are separated by about 450 days.

Eq.~\ref{DB} indicates $|B_\parallel| \sim 17$ mG for the transition between
groups I and II and $|B_\parallel| \sim 3$ mG for the transition between
groups II and III.  Within this model, these estimates of the magnetic field
do not depend on the times over which the transitions occurred.  These
fields are three or more orders of magnitude greater than typical estimated
interstellar fields, and indicate that FRB 121102 is behind or immersed in
an extraordinary region.

Assuming that the entire DM of the dense region, DM$_{dr}$, varies when its
RM varies, 
\begin{equation}
	\label{DMdr}
	\text{DM}_{dr} \sim {|\text{RM}| \over |\Delta \text{RM}|}
	|\Delta \text{DM}| \sim 10\,\text{pc/cm}^3.
\end{equation}
This is only a rough estimate, but is consistent with the attribution of
most of the DM to interstellar and intergalactic media, even though nearly
all the RM must be attributed to the dense region.
\section{Numerical Estimates}
If the moving plasma that is the source of the changing DM and RM has a
speed (that might be the phase speed of an ionization or recombination front
rather than a material speed) $v$, then a change in properties over a time
$\Delta t$ corresponds to a length scale $v \Delta t$.  The characteristic
electron density
\begin{equation}
	\label{ne}
	n_e \sim {\Delta \text{DM} \over v \Delta t}.
\end{equation}

If the magnetic stress is in equipartition with the gas pressure $P = 2 n_e
k_B T$ (assuming equal ion and electron temperatures and a statistically
isotropic field), there is a characteristic temperature $T$:
\begin{equation}
	\label{P}
	2 n_e k_B T = P = 3{B_\parallel^2 \over 8 \pi}.
\end{equation}
Combining Eqs.~\ref{DB}, \ref{ne} and \ref{P},
\begin{equation}
	\label{T}
	k_B T = 2 \times 10^{-13} {|\Delta \text{RM}|^2 \over |\Delta
	\text{DM}|^3} {\Delta t \over \text{1 s}} v_7\ \text{eV},
\end{equation}
where RM and DM are in their usual units (radians/m$^2$ and pc/cm$^3$) and
$v_7 \equiv v/(10^7\text{cm/s})$.

Alternatively, the magnetic stress could be equated to the dynamic pressure
$n_e \mu v^2$, where $\mu$ is the molecular weight per electron, taken as
$2 \times 10^{-24}\,$g:
\begin{equation}
	\label{Q}
	n_e \mu v^2 = 3{B_\parallel^2 \over 8 \pi}.
\end{equation}
Combining Eqs.~\ref{DB}, \ref{ne} and \ref{Q},
\begin{equation}
	\label{v}
	v = 3 \times 10^{-8} {|\Delta\text{RM}|^2 \over |\Delta\text{DM}|^3}
	{\Delta t \over \text{1 s}}\ \text{cm/s}.
\end{equation}

The inferred values of the parameters are shown in the Table.  The
assumptions, particularly those of equipartition, are uncertain, but the
inferred orders of magnitude of $k_B T$ and $v$ may be plausible in regions
of high energy density.

\begin{table}
	\centering
	\begin{tabular}{|l|cc|}
		\hline
		 Epoch & I$\to$II & II$\to$III \\
		\hline
		 $\Delta$RM (rad/m$^2$) & $-15000$ & $-4000$ \\
		 $\Delta$DM (pc/cm$^3$) & $+1.1$ & $+1.6$ \\
		 $|B_\parallel|$ (mG) & 17 & 3 \\
		 $\Delta t$ (s) & $1.5 \times 10^7$ & $4 \times 10^7$ \\
		 $\ell = v \Delta t$ (cm) & $1.5 \times 10^{14}\,v_7$ &
		 $4 \times 10^7\,v_7$ \\
		 $n_e$ (cm$^{-3}$) & $2 \times 10^4\,v_7^{-1}$ &
		 $1 \times 10^4\,v_7^{-1}$ \\
		 $L$ (cm) & $10^{15} v_7$ & $10^{16} v_7$ \\
		 $k_B T$ (eV) & 500 $v_7$ & 30 $v_7$ \\
		 $v_7$ & 7 & 0.5 \\
		 \hline
	\end{tabular}
	\caption{\label{table1} The parameters of the plasma responsible for
	changes in DM and RM of FRB 121102 between Epochs I (bursts 5 and 6)
	and II (bursts 7-15) and between Epochs II and III (bursts 16--20)
	of \citet{H20}, estimated from Eqs.~\ref{DB} and \ref{ne}.  $k_B T$
	is estimated from Eq.~\ref{T}, assuming magnetic equipartition with
	the gas pressure with $v$ as a parameter, and $v$ is estimated from
	Eq.~\ref{v}, assuming magnetic equipartition with dynamic pressure.}
\end{table}

An additional constraint may be imposed on $n_e$ by the observation that the
dispersion index $\alpha = -1.997 \pm 0.015$ \citep{S16}.  Most of the DM is
produced in low density interstellar and intergalactic plasma whose
contribution to any deviation from $\alpha = -2$ is infinitesimal.  The
portion of the DM attributable to the dense magnetized region
(Eq.~\ref{DMdr}) is $\sim 2\%$ of the total DM, so in the dense region the
emprical bound is $\Delta \alpha_{dr} \sim (-\alpha - 2)/0.02 \lesssim 0.6$,
implying \citep{K14,K15}
\begin{equation}
	n_e \approx {4 \over 3} {\omega^2 m_e \over 4 \pi e^2} \Delta
	\alpha_{dr} \lesssim 2 \times 10^{10}\ \text{cm}^{-3}.
\end{equation}
This is not restrictive.

The electron density inferred for FRB 121102 implies very short ($\sim
10^9\,$s, unless the temperature is very high) recombination times.  This
may suggest very young ages, but might also be explained by an intense
ionizing ultraviolet flux.
\section{Implications}
FRB 121102 is behind or immersed in an extraordinarily dense and strongly
magnetized region, plausibly associated with the FRB source or with the
event that made it.  The inferred $n_e$ and $|\langle B_\parallel
\rangle_{n_e}|$ are much greater than those found in other interstellar
plasmas.  The data are consistent with any location of this region on the
line of sight between the FRB source and the observer, but it is natural to
associate it with the immediate environment of the source of FRB 121102.
Combining Eqs.~\ref{DMdr} and \ref{ne} suggests a length scale
\begin{equation}
	\label{L}
	L \sim {\text{DM}_{dr} \over n_e} \sim {|\text{RM}| \over
	|\Delta\text{RM}|} v \Delta t \sim 10^{15} v_7\ \text{cm}.
\end{equation}

The identification of FRB 200428 with SGR 1935$+$2154, associated with the
supernova remnant (SNR) G57.2$+$0.8 \citep{KSGR18}, suggests by analogy that
the dense region on the line of sight to FRB 121102 may be a very young,
dense and strongly magnetized remnant of the event that made its source.
However, the magnetoionic environment of SGR 1935$+$2154 is very different,
with a modest RM plausibly attributed to the interstellar medium.   

This explanation of the RM and DM of FRB 121102 is not entirely
satisfactory.  The expansion of a SNR generally produces a decreasing DM
\citep{MM18,PG18}, in contradiction to the observed \citep{H20,O20}
increase, although increasing ionization could increase DM.  More complex
models \citep{ZZWW} may have more complex behavior.  A spherically symmetric
remnant cannot have a magnetic field at all, but a remnant might be
approximately symmetric on large scales while being asymmetric (turbulent or
otherwise structured) on smaller scales.

The changing DM and RM of FRB 121102 could be attributed to either a change
in the magnetic field and density of a nebular filament on the line of sight
or to the motion of such a filament across the line of sight, as suggested
for PSR J1745$-$2900 by \citet{P18}.  The observed reduction in |RM|,
combined with an increase in DM, implies the movement into the line of sight
of comparatively dense plasma whose $B_\parallel$ is opposite to that
dominant along the total line of sight, or an increase in the integrated
column density of such an opposed region.

It is natural to compare the inferred parameters with those of the
best-studied SNR, the Crab Nebula.  Its magnetic field \citep{BK90,RGB12} is
less than that inferred for FRB 121102.  However, direct comparison is 
inappropriate because the magnetic field of the Crab nebula is inferred for
the low density synchrotron-radiating volume that must be a negligible
contributor to both DM and RM; it is not possible to measure the field in
the high density filaments directly.  Further, both the Crab Nebula and its
central neutron star differ from those that produce FRB 200428, and very
plausibly other FRB.  FRB 200428/SGR 1935$+$2154 has a spin rate about 100
times less than that of the Crab pulsar and a magnetic moment about 60 times
greater.  The age of FRB 121102 may be much less than that of the Crab
Nebula and its pulsar; it must be at least eight years, but is otherwise
empirically unconstrained.
\section{Discussion}
It is remarkable that the RM and inferred $\langle B_\parallel
\rangle_{n_e}$ of FRB 121102 are orders of magnitude greater than those of
any other FRB \citep{WWY}\footnote{\citet{WWY} use Eq.~\ref{B} rather than
Eq.~\ref{DB} to infer $\langle B_\parallel \rangle_{n_e}$, leading to an
underestimate if most of the RM but little of the DM are contributed by a
dense strongly magnetized region.  For FRB 121102, most of whose RM must be
produced in such a region (because the evidence of other FRB shows that
interstellar and intergalactic plasma contribute much less RM), Eq.~\ref{B}
may underestimate $\langle B_\parallel \rangle_{n_e}$ by a factor $\sim
(\Delta \text{RM}/\text{RM})(\text{DM}/\Delta \text{DM}) \sim 50$.  The
error is much less if most of the RM is produced in low density Galactic and
host galactic interstellar media, as it may be for other FRB.}.  If the
sources of the DM are qualitatively similar in FRB 121102 and other FRB, as
would be expected if they are gradually expanding and dissipating SNR with a
continuous distribution of parameters (such as age, electron density and
magnetic field), a smooth and continuous distribution of RM would be
expected, perhaps a power law.  This is inconsistent with the extraordinary
and unique RM and inferred $\langle B_\parallel \rangle_{n_e}$ of FRB
121102; its environment differs qualitatively from those of other FRB, and
its nature and origin may also differ.

The source of FRB 200428 shows an analogously anomalous distribution of
burst strengths, with FRB 200428 orders of magnitude more intense than
subsequent bursts from this source.  This may be explained by the
collimation of radiation emitted by relativistic charges into a narrow beam
that may jitter or wander in direction \citep{K20}.  No such explanation is
apparent for the distribution of RM among FRB sources.

\citet{H20} compared the observed properties of FRB 121102 to the SNR models
of \citet{MM18} and \citet{PG18}, and suggested birthdates in the range
2000--2010.  The rapid decay of RM (30\% in three years), if interpreted as
exponential decay, would also suggest a characteristic age of about 10
years.  No trend in the rate and strengths of outbursts of FRB 121102, that
might be expected to decay as a young source ages, has been qualitatively
evident in eight years of observation, but insufficient data exist to bound
any trends quantitatively.  The \emph{increase} in DM argues against a
rapidly dissipating (and therefore very young) SNR.  This increase, and the
fact that FRB 121102 has been observed over a time scale $\gtrsim \tau
\equiv |\text{RM}|/|d\text{RM}/dt|$ without evident decay, suggests that its
environment is fluctuating chaotically rather than systematically evolving
and that its age and life expectancy may be $\gg \tau$.  As discussed in the
Appendix, characteristic time scales of variation are only lower bounds on
actual ages.

If a repeating FRB had a measured period and period derivative, these would
constrain its age, although the possibility of a neutron star born recently
with a long spindown age could not be excluded---spindown age is only an
upper bound on the actual age.  Because FRB are episodic emitters, a birth
date could only be established statistically from its absence in surveys
prior to some date; CHIME/FRB might make this possible.

The only other astronomical objects with such extraordinarily large (and
rapidly varying) RM are PSR J1745$-$2900, the similarity of whose properties
to those of FRB was suggested by \citet{P18}, and the Galactic center black
hole Sgr A$^*$.  This PSR has a projected separation from Sgr A$^*$ of only
0.1\,pc, so the line of sight to PSR J1745$-$2900 may pass through a region
filled with plasma accreting onto the massive black hole.  It is
unsurprising that this region has unique properties, and \citet{K19}
suggested that FRB are emitted by jets produced by intermediate mass black
holes.  That cannot explain FRB 200428, known to be emitted by a SGR (and
that has an unremarkable RM, plausibly attributed to interstellar plasma),
but is consistent with the hypothesis that there are at least two,
qualitatively different, kinds of objects that produce FRB whose
phenomenologies are not obviously distinct.

It is difficult to explain the rapid variation of the RM of PSR J1745$-$2900
as the result of a path through an accretion flow onto Sgr A$^*$.  A
characteristic time scale $\sim 6\,$y (10\% change in 7 months)
\citep{MSH18} at a projected distance from Sgr A$^*$ of 0.1 pc would suggest
a velocity $\sim 1.5 \times 10^9\,$cm/s, about 50 times the escape or
orbital velocity of the $4 \times 10^6 M_\odot$ black hole in Sgr A$^*$.
For a smaller and more plausible $v$ the scale $L$ (Eq.~\ref{L}) of the
birefringent plasma is $\ll 0.1\,$pc.  Alternative explanations include
propagation through a very fast and unslowed supernova shell and varying
refraction, in which the high velocity is not a material velocity but a
displacement velocity of the propagation path as a result of refraction by
more slowly moving material.  Application of this hypothesis to FRB 121102
would not invalidate the inference of extraordinary $n_e$ and $\langle
B_\parallel \rangle_{n_e}$, but would emphasize that $v$ is uncertain.

Only upper limits to |$\Delta$DM| of PSR 1745$-$2900 exist, so it is not
possible to estimate $\langle B_\parallel \rangle_{n_e}$.  However, these
upper limits are comparable to the measured |$\Delta$DM| of FRB 121102 and
the $\Delta\,$RM are of the same order of magnitude, so $|\langle B_\parallel
\rangle_{n_e}| \sim 10\,$mG is at least plausible.

If varying refraction explains the rapid variation of the RM of FRB 121102,
we cannot be on a caustic.  The converging rays of a caustic, while having
the same optical path (foci occur at extrema of the optical paths), would
traverse paths of differing $B_\parallel$, averaging out their linear
polarization.

The large and rapidly varying RM of SGR/PSR J1745$-$2900 may be unrelated to
its proximity to Sgr A$^*$ (a separation of $0.1\,\text{pc} \gg v \Delta t$
for likely $v$), but may rather be associated with the environment of the
SGR/PSR on scales $\ll 0.1\,$pc.  PSR J1745$-$2900 and FRB 121102 may be
members of a distinct class of object characterized by the behavior of their
RM.  If so, then the FRB activity of FRB 121102 suggests that PSR
J1745$-$2900 might also be a source of FRB, a speculation supported by the
discovery that another SGR/PSR with similar parameters (1935$+$2154) made
FRB 200428.
\section*{Appendix}
It has been suggested \citep{G93} on the basis of a ``Copernican Principle''
(of much older origin) that if a phenomenon has been observed for a period
$t$, the probability that its lifetime is $\gtrsim T$ is ${\cal O}(t/T)$.
This has been used to argue that the expected lifetime of humanity
(estimated to be $\sim 300,000$ years old, depending on the definition of
when our ancestors became human) is unlikely to be more than a few million
years, that the expected lifetime of the present American constitutional
government, established in 1788, is unlikely to be more than a few thousand
years, {\it etc.\/}

This argument depends on an assumed uniform prior probability of discovery
through the lifetime of the phenomenon.  This can be shown to be wrong, at
least in some applications.  For example, the rapid variation of some AGN
might suggest ages of no more than a few years, but we are confident, on the
basis of the physics of black hole formation and accretion and observation
of their accompanying double radio sources, that their ages are
$10^6$--$10^{10}$ years.  The variation of the RM of PSR J1745$-$2900
suggests an age $\approx 6\,$y, but the absence of a neutrino signal in
Kamiokande and other neutrino observatories from its formation implies
an age $\ge 37\,$y, and the properties of SNR G57.2$+$0.8 in which it is
embedded indicate an age of thousands or tens of thousands of years
\citep{KSGR18}.  Hence, while the rapid variation of the RM of FRB 121102
may suggest on physical grounds a very young SNR, its age cannot be
estimated by a statistical argument. 
\section*{Data Availability}
This theoretical study did not obtain any new data.

\label{lastpage} 

\begin{thebibliography}{99}
	\bibitem[\protect\citeauthoryear{Bietenholz \& Kronberg}{1990}]
		{BK90} Bietenholz, M. F. \& Kronberg, P. P. 1990 \apjl\
		357, L13.
	\bibitem[\protect\citeauthoryear{Gott}{1993}]{G93} Gott, J. R. 1993
		\nat\ 363, 315.
	\bibitem[\protect\citeauthoryear{Hilmarsson {\it et al.\/}}{2020}]
		{H20} Hilmarsson, G.~H., Michilli, D., Spitler, L.~G. {\it
		et al.\/} 2020 arXiv:2009.12135.
	\bibitem[\protect\citeauthoryear{Katz}{2014}]{K14} Katz, J. I. 2014
		\apj\ 788, 34. 
	\bibitem[\protect\citeauthoryear{Katz}{2015}]{K15} Katz, J. I. 2015
		arXiv:1505.06220.
	\bibitem[\protect\citeauthoryear{Katz}{2018}]{K18} Katz, J. I. 2018
		Prog. Nucl. Part. Phys. 103, 1 arXiv:1804.09092.
	\bibitem[\protect\citeauthoryear{Katz}{2019}]{K19} Katz, J. I. 2019
		\mnras\ 487, 491 arXiv:1811.10755.
	\bibitem[\protect\citeauthoryear{Katz}{2020}]{K20} Katz, J. I. 2020
		\mnras\ 499, 2319 arXiv:2006.03468.
	\bibitem[\protect\citeauthoryear{Kothes {\it et al.\/}}{2018}]
		{KSGR18} Kothes, R., Sun, X., Gaensler, B. \& Reich, W. 2018
		\apj\ 852, 54.
	\bibitem[\protect\citeauthoryear{Margalit \& Metzger}{2018}]{MM18}
		Margalit, B. \& Metzger, B. D. 2018 \apjl\ 868, L4.
	\bibitem[\protect\citeauthoryear{Michilli, Seymour \& Hessels}
		{2018}]{MSH18} Michilli, D., Seymour, A. \& Hessels,
		J.~W.~T. 2018 \nat\ 553, 182.
	\bibitem[\protect\citeauthoryear{Oostrum {\it et al.\/}}{2020}]{O20}
		Oostrum, L. C., Maan, Y., van Leeuwen, J. {\it et al.\/}
		2020 \aap\ 635, 61.
	\bibitem[\protect\citeauthoryear{Pearlman {\it et al.\/}}{2018}]
		{P18} Pearlman, A. B., Majid, W. A., Prince, T. A. {\it et
		al.\/} 2018 \apj\ 866, 160.
	\bibitem[\protect\citeauthoryear{Piro \& Gaensler}{2018}]{PG18}
		Piro, A. L. \& Gaensler, B. M. 2018 \apj\ 861, 150.
	\bibitem[\protect\citeauthoryear{Reynolds, Gaensler \& Bocchino}
		{2012}]{RGB12} Reynolds, S. P., Gaensler, B. M. \& Bocchino,
		F. 2012 Sp.~Sci.~Rev. 166, 231.
	\bibitem[\protect\citeauthoryear{Scholz {\it et al.\/}}{2016}]{S16}
		Scholz, P., Spitler, L.~G., Hessels, J.~W.~T. {\it et al.\/}
		2016 \apj\ 833, 177.
	\bibitem[\protect\citeauthoryear{Wang {\it et al.\/}}{2020}]{WWY}
		Wang, W.-Y., Zhang, B., Chen, X. \& Xu, R. 2020 \mnras\
		499, 355.
	\bibitem[\protect\citeauthoryear{Zhao {\it et al.\/}}{2020}]{ZZWW}
		Zhao, Z. Y., Zhang, G. Q., Wang, Y. Y. \& Wang, F. Y. 2020
		arXiv:2010.10702.
\end{thebibliography}
\end{document}